\documentclass{iaus}
\usepackage{graphicx}

  \newcommand {\vmi}   {$v_{\mathrm {micro}}$}

  \newcommand {\Tef}   {$T_{\mathrm {eff}}$}
  
  \newcommand {\lgg}   {$\mathrm \log ~g$}
  \newcommand {\vsini} {$\mathrm~{v\sin i}$~}
  \newcommand {\kms}   {${\mathrm {km~s^{-1}}}$}
  \newcommand {\dsct}	{$\delta$~Sct~}
  \newcommand {\dsctM}	{\delta~Sct}

\title[Chemical composition of $\delta$ Scuti] %% give here short title %%
{About the  chemical composition  of $\delta$ Scuti -
the prototype of the class of pulsating variables}

\author[Yushchenko et al.]   %% give here short author list %%
  {A.V.Yushchenko$^{1, 2}$
   V.F. Gopka$^2$,
   Chulhee Kim$^3$,
   F.A. Musaev$^{4, 5, 6}$,
   Y.W. Kang$^1$}
\affiliation{
  $^1$ Astrophysical Research Center
         for the Structure and Evolution of the Cosmos (ARCSEC),
         Sejong University, Seoul, 143-747,  (South) Korea
         \break email:yua@odessa.net
         \\[\affilskip]
  $^2$ Odessa Astronomical observatory, Odessa National University,
         Park Shevchenko, Odessa, 65014, Ukraine
         \break email: gopkavera@mail.ru \\[\affilskip]
  $^3$ Department of Earth Science Education,
         Chonbuk National University,
         Chonju 561-756, (South) Korea \break email: chkim@astro.chonbuk.ac.kr
         \\[\affilskip]
  $^4$ Special Astrophysical observatory of the Russian Academy of Sciences,
         Nizhnij Arkhyz, Zelenchuk, Karachaevo-Cherkesiya, 369167,
         Russia \break email: faig@sao.ru    \\[\affilskip]
  $^5$ The International Centre for Astronomical, Medical and Ecological
         Research of the Russian Academy of Sciences and
         the National Academy of Sciences of Ukraine,
          \\[\affilskip]
  $^6$ Shamakhy Astrophysical Observatory,  NAS  of Azerbaijan, Yusif
         Mamedaliev, Shamakhy, Azerbaijan  \\[\affilskip]
    }
\pubyear{2004}
\volume{224}  %% insert here IAU Symposium No.
\pagerange{119--126}
\date{?? and in revised form ??}
\jname{The A-Star Puzzle}
\editors{J.\,Zverko, W.W.\,Weiss, J.\,\v{Z}i\v{z}\v{n}ovsk\'{y}, \& S.J.\,Adelman, eds.}
\begin{document}
\maketitle
\begin{abstract}
  We  present  chemical  abundances in the photosphere of $\delta$ Scuti --
  the prototype of the class of pulsating variables --
    determined  from  the  analysis  of a spectrum
 obtained at Terskol observatory 2 meter telescope
 with resolution $R=52,000$, signal to noise ratio
  250.  VLT and IUE spectra were used also .
  Abundance pattern of \dsct consists of 49 chemical elements.
  The abundances of
  Be, P, Ge, Nb, Mo, Ru, Er, Tb, Dy, Tm, Yb, Lu,  Hf, Ta,  Os,  Pt,  Th
  were not investigated previously.
  The lines of third spectra of Pr and Nd also are investigated
  for the first time.
  The abundances of heavy elements show the overabundances with respect
  to the Sun up to 1~dex.
  The abundance pattern of \dsct is similar to that of  Am-Fm stars.
   \end{abstract}
            \firstsection % if your document starts with a section,

   \section{Introduction}

     \dsct  type  are  pulsating variables
   located in the lower part
  of  the  instability strip.
     This type stars form the second most numerous group of pulsators in
  the  Galaxy after the pulsating white dwarfs (Solano \& Fernley 1997).
  New catalog (Rodriges \& Breger 2001) has observational information
  available up to January 2000  about 636 stars of this type .
    The  majority of \dsct type stars belongs  to Population I.

  The chemical composition of \dsct was investigated by
  Russell (1995), Rachkovskaya (2000).
  The most detailed abundance pattern  was published by
  Erspamer \& North (2003). They found the abundances of 30 elements.

    The goal of this paper is to derive detailed abundances of chemical
   elements in the atmosphere of   \dsct  -- prototype of the
 numerous class of pulsating variables.

\section{Observations and data reduction}

  High  resolution spectrum of \dsct was  obtained
  using  a coude-echelle  spectrometer (Musaev  et al.\ 1999)
  mounted on the 2-m ``Zeiss'' telescope at the Peak Terskol Observatory
  located  near Mt. Elbrus  (Northern  Caucasus,  Russia)
  3,124 m above sea level.

\begin{table} % \def~{\hphantom{0}}
  \begin{center}
 \caption{Observations}
  \begin{tabular}{llcrcrcl}
\hline
telescope &        Spectral range      &  Resolving   & S/N & JD &Exposure \\
          &        (\AA\AA)            &    power     &     &    &  (sec)   \\
\hline
  Terskol 2 m         &3610--10270& R=52000  & 250  &2452422.4734 & 1200         \\ % 27.05.02  23:15
  VLT 8 m             &3860--4980 & R=80000  &$>$300&2452008.4119 &   40         \\ %   52008.41102
                      &6805--8540 & R=80000  &$>$300&2452008.4119 &   40         \\ %        .41107
                      &8662--9194 & R=80000  &$>$300&2452008.4119 &   40         \\ %        .41107
  IUE 0.4 m           &1850--3349 &  0.2~\AA & 20   &2444788.1094 & 2099         \\ % 1981-07-02 14:11:54
                      &1850--3349 &  0.2~\AA & 20   &2448115.1387 &  720         \\ % 1990-08-11 15:07:19
                      &1850--3349 &  0.2~\AA & 20   &2448120.1593 &  720         \\ % 1990-08-16 15:37:31
\hline
  \end{tabular}
 \end{center}
 \label{ObservT}
\end{table}

  We  used   the spectrograph in the mode   with a resolving
  power  of  $R=52,000$.
  The  observed  wavelength  range,
  $\lambda\lambda~$3610--10270 \AA\AA,  was  covered  by  86  echelle
  orders. In the observed spectrum, there are gaps between the orders in the
  wavelength  region  of $\lambda~\geq~6705$~\AA\ and the width of each
  gap  increases from 0.5~\AA\ to  69~\AA\ as the wavelength increases.
  The signal-to-noise reaches  250 or more in the red part of the  spectrum.

  The first-stage  data  processing (background  subtraction, echelle vector
  extraction  from  the echelle-images,  and  wavelength calibration) was
  done	 using	the  latest  version  of  PC-based DECH software
  (Galazutdinov  1992).
  For  other  processes  including  continuum
  placement,  we  use URAN software (Yushchenko 1998).
  The location of the continuum was determined
  taking into account the calculated  spectrum.

  The strongest lines of many chemical elements can be observed
  in the ultravilet spectral region. To detect these lines
  we used three IUE spectra of \dsct from INES archive --
  the spectra  LWR10992HL, LWP18563HL, LWP18600HL.
  The wavelength coverage of these spectra is from 1850 to 3349~\AA,
  spectral resolution is near 0.2 \AA. The signal to noise ratio
  is sufficient for line identification and in some cases
  for deriving abundances  with errors near 0.2-0.3 dex.

  The spectra of \dsct from VLT archive were also used.
  The spectral
  resolwing power of this spectra are 80,000, S/N ratio more than 300,
  wavelength coverage 3860--4980 and 8661--9194~\AA.

  Short information about all used spectra can be found in
  Table~1.%\ref{ObservT}.
  ~For abundance determinations the  spectrum of Terskol observatory
  and IUE spectra were used, VLT spectrum was used for identification
  of faint line.

\section{Atmosphere parameters}

  The information about previous determinations of effective temperature,
  surface gravity, rotation and microturbulence of \dsct can be found in
  Table~\ref{Parameters}.
  We tried to find our values of atmospheric parameters.
  The values obtained from Geneva photometry and
  from the depth ratios of the iron  lines, based on Kovtyukh \& Gorlova (2000)
  method are listed at the end of the table.

\begin{table}
\caption { Atmospheric parameters of \dsct  }
\begin{tabular}{lcllc}
\hline
			      &  \Tef	   & \lgg &\vsini & \vmi  \\
  \multicolumn{1}{c}{}	  &
  \multicolumn{1}{c}{(K)} &
  \multicolumn{1}{c}{}	  &
  \multicolumn{2}{c}{(\kms)} \\
 \hline
  Philip \& Relyea 1979   & 7300  &	 &	 &	 \\% uvbyBeta
  Moon \& Dvoretsky 1985  & 7200  &	 &	 &	 \\% uvbyBeta
  Lester et al. 1986	  & 7100  &	 &	 &	 \\% uvbyBeta
			  & 7000  &	 &	 &	 \\%
  Balona  1994		  & 7267  &	 &	 &	 \\%
  Russel  1995		  & 7200  & 3.71 & 39	 &  2.5  \\% 1995 ApJ 451 747 Russell, S.C.
  Solano \& Fernley 1997  & 7000  &	 & 30.1  &	 \\% Solano, E. \& Fernley, J. 1997 A&AS 122 131
			  & 6900  &	 &	 &	 \\%
  Rachkovskaya 2000	  & 7000  & 3.1  & 32	 &  5.4  \\% Rachkovskaya, T.M. 2000 ARep 44 227
			  &	  &	 & 30	 &	 \\%
  Erspamer \& North 2003  & 6776  & 3.47 & 25.51 &  2.8  \\% Erspamer, D.; North, P.  2003 A&A 398 1121
  Geneva system 	  & 6772  & 3.45 &	 &	 \\%
  Iron lines depth ratios & 7064  &      &       &       \\
  Adopted values          & 7000  & 3.5  & 25.5  &  3.8  \\%
 \hline
 \end{tabular}
 \label{Parameters}
 \end{table}

  We adopted the values of effective temperature and surface gravity
  \Tef=7000~K, \lgg=3.5.
  This parameters, Erspamer \& North (2003) abundances
  and the values  microturbulence and rotation velocities were used
  for initial calculation of synthetis spectrum in the whole observed
  region.   This synthetic spectrum was used for identification of
  clean iron lines in the spectrum. For these lines equivalent
  widths were found and the values of parameters were tested.
  It was necessary to change the microturbulence velocity
  to the value \vmi=3.8~\kms.

  The set of parameters (\Tef=7000~K, \lgg=3.5, \vmi=3.8~\kms) and
  Erspamer \& North (2003) abundances were used to produce
  individual atmosphere model using Kurucz (1995) ATLAS12 code.
  The dependancies of
  iron abundances on the equivalent widths  and on the excitation levels
  of individual lines iron lines in the spectrum of \dsct
  for our model show zero correlation.
  This model was used for abundance calculations.

 \section{Methodics}

 Differential  spectrum  synthesis  method is  used
 for all  elements, except iron.
 For each line, we tried to find its counterpart in the solar
 spectrum atlas of Delbouille et al. (1973).
 Grevess \& Sauval (1999) solar photosphere model was used.
 This procedure frees us from
 uncertainties	connected  with oscillator  strengths  of spectral lines.
 The URAN code (Yushchenko 1998) and SYNTHE spectrum synthesis program
 (Kurucz 1995)
 are used to approximate the observed spectrum by the synthetic one.

 A  synthetic spectrum  of \dsct for the whole wavelength range
 helps us to identify  spectral  lines.
 Are included atomic and molecular lines
 from  Kurucz (1995) as well as Morton (2000),
 Biemont  et  al. (2002)  and partially
 from the VALD database (Piskunov  et  al.\ 1995).

 Hyperfine structure and isotopic splitting are taken into account
 for Sc, V, Mn, Cu, Ba, Eu. The splitting data for Ba are taken from
 Francois (1996) and for other elements from Kurucz (1995).
 It should be noted that for all elements except Li, S, K, Ir, Th, U
 we found  counterparts in the solar spectrum, so that the differential
 abundances are not strongly influenced by splitting effects.
 Holweger's partition function for thorium is used (Morell et al. 1992).

\section{The abundance pattern of $\delta$ Scuti}

 In Tables~\ref{meanT} and ~\ref{mIUE} the
 mean elemental abundances in the atmosphere of \dsct are given.
 This tables contains  data, obtained from the spectrum of
 Terskol observatory and IUE specta respectively.

 \begin{figure}
 \centerline{ \includegraphics[width=4.0in]{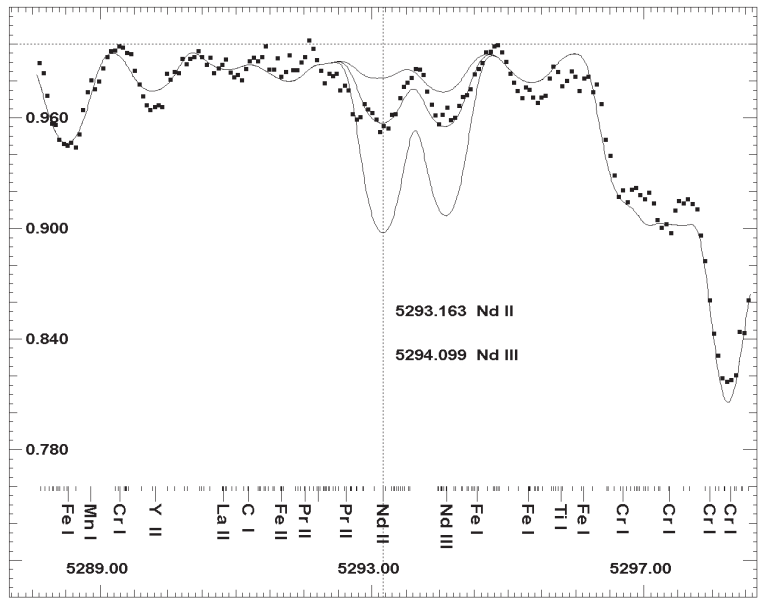} }
 \caption[]
 { The observed spectrum of  \dsct (squares) and
 the synthetic spectra (solid lines)
 calculated
 with our final abundances.
 The axes are the wavelength in angstroms and relative fluxes.
 The positions of the spectral lines taken into account
 in the calculations are marked in the bottom part of the figure.
 For some of the strong lines the identification are given.
 The position of the  Nd~II~$\lambda$~5293.163~\AA~  line is marked by a
 vertical dotted line.
 The different synthetic spectra correspond to
 a Nd abundance lower or higher by 0.5~dex with respect to the abundance
 obtained from the optimum value.
 Nd~II~$\lambda$~5293.163~\AA~ and Nd~III~$\lambda$~5294.099~\AA~ lines
 show the
 example of the lines of the second and the third spectra of neodymium.
 The approximation of both spectra can be made with one value of abundance.
 }
 \label{fig-Nd}
 \end{figure}

 The difference in temperature between \dsct
 and that of the Sun is quit large and we were not able to find the
 counterparts in the solar spectrum for all investigated lines.
 That is why in Table~\ref{meanT}
 we give both relative and absolute abundances of the chemical elements
 in the atmosphere of \dsct.

\begin{figure}
  \centerline{ \includegraphics[width=5.0in]{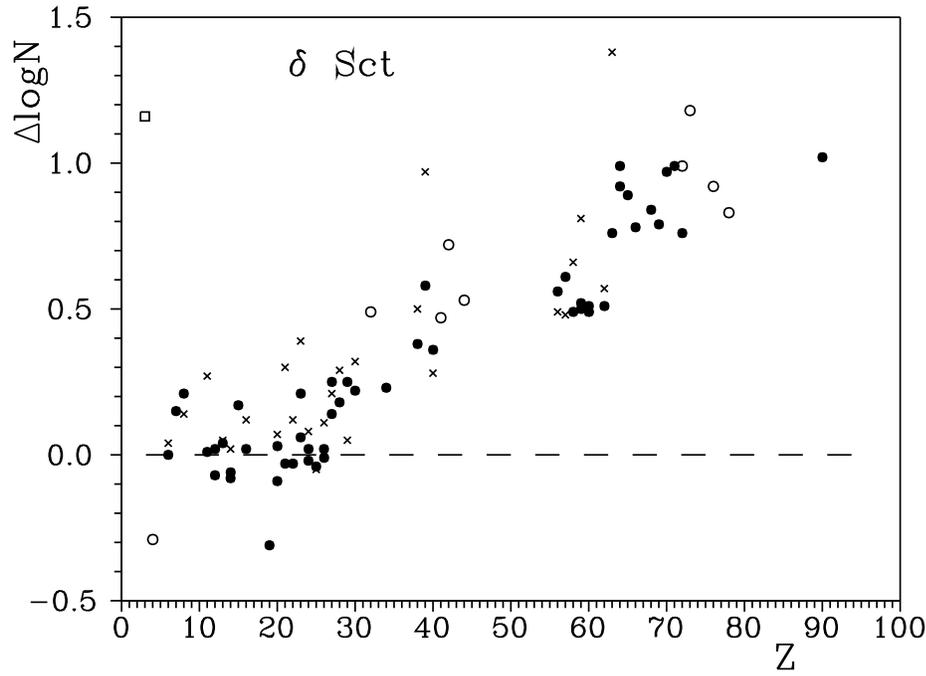} }
  \caption{
     The abundances of chemical elements and ions in the atmosphere
	 of \dsct  with respect to their
	 abundances in the solar atmosphere.
         Filled circles -- Terskol observatory observations.
	 Open circles -- IUE data. Crosses -- Erspamer \& North (2003).
         Open square -- lithium (Russel 1995)
     }\label{fig:wave3}
\label{meanAb}
\end{figure}

 The first two columns are the atomic number and identification of
 investigated species. The next two are the number of lines with
 counterparts in the solar spectrum and the relative abundance, obtained from
 these lines, with the last figures of errors in brackets.
 The fifth column is the total number of lines of given element or ion.
 Two subsequent triplets of colunms are the mean abundances in the atmosphere
 of \dsct for three sets or atmospheric parameters: the best values,
 and effective temperature and surface gravity shifted by -0.2~dex and
 +100~K respectively. There are absolute values of abundances in the second
 triplet and relative, with respect to Grevesse \& Sauval (1998),
 in the first.

 Table~\ref{mIUE} contains the similar values
 for lines, investigated using IUE spectra. Only absolute abundances
 and relative values with respect to Grevesse \& Sauval (1998)
 are shown in this table.
 Fig.~\ref{meanAb} illustrates the relative abundances in the atmospere of
 \dsct.

 The analysis of the tables shows the relative abundances calculated with
 respect to the absolute solar photosphere abundances are very close to the
 direct differential abundances.

\section{Conclusions} \label{sec:concl}
\begin{itemize}
\item  The abundance pattern of \dsct consists of
   49 chemical elements. The lines of eight of them are found only in
   ultraviolet spectrum.
\item  The abundances of Be, P, Ge, Nb, Mo, Ru, Er, Tb, Dy, Tm, Yb, Lu,
   Hf, Ta,  Os,  Pt,  Th
   were not investigated previously.
\item  The lines of the third spectra of Pr and Nd are observed.
   The values of abundances of these elements
   obtained from the lines of second and third spectra are equal.
\item  The abundances of heavy elements shows the overabundances with respect
   to the Sun up to 1~dex.
\item  The abundance pattern of \dsct is similar to that of  Am-Fm stars.
\end{itemize}

\section {Acknowledgments}

  We  would like to thank to L. Delbouille and G. Roland for sending us
  the  Liege  Solar  Atlas.
  We used data INES data from the IUE satellite,
  the data  from  NASA ADS,  SIMBAD,  CADC,  VALD,  NIST, and DREAM
  databases,
  the  data from the UVES Paranal Observatory Project
  (ESO DDT Program ID 266.D-5655)
  and  we  thank the  teams  and  administrations of these
  projects.

  Work by AY and YK was supported by the Astrophysical Research Center
  for the Structure and Evolution of the Cosmos (ARCSEC) of Korea
  Science and Engineering Foundation (KOSEF) through the Science
  Research Center (SRC) program. Work by VG and CK was supported
  by research funds of  Chonbuk National University, Korea.

\begin{table}
\caption { Mean abundance of chemical elements in the atmosphere of \dsct.
	   Terskol observatory observations}
\begin{tabular}{rl c rl c rlll c lll }
\hline
 & &&\multicolumn{2}{c}{$\Delta$logN$_{\dsctM ~-~ \odot}$ }&&
     \multicolumn{4}{c}{$\Delta$logN$_{\dsctM ~-~ (Grevesse~\&~Sauval~1998)}$}&&
     \multicolumn{3}{c}{	logN$_{\dsctM}$ 			     }\\	      %
 \cline{4-5}  \cline{7-10}  \cline{12-14}						      %
  Z & Ident. && n& $*-\odot$								      %     Li	      1
			   && n&$*$--GS98&\lgg-0.2&\Tef+100  && logN   &\lgg-0.2 &\Tef+100 \\ %     Be	      2   Be
\hline											      %
  6 &C~~I     &&  7&~0.00(06)&& 14&~0.00(12)&-0.02(09)&~0.01(09)&&8.52(12)& 8.50(09)& 8.53(09)\\ %  6 &C~~I	 3
  7 &N~~I     &&  3&~0.04(11)&&  4&~0.15(06)&~0.04(11)&~0.02(11)&&8.07(06)& 7.96(11)& 7.94(11)\\ %  7 &N~~I	 4
  8 &O~~I     &&  4&~0.23(18)&&  5&~0.21(17)&~0.15(17)&~0.17(17)&&9.04(17)& 8.98(17)& 9.00(17)\\ %  8 &O~~I	 5
 11 &Na~I     &&  3&~0.07(12)&&  6&~0.01(02)&~0.04(07)&~0.04(15)&&6.34(02)& 6.37(07)& 6.37(15)\\ % 11 &Na~I	 6
 12 &Mg~I     &&  2&-0.04(15)&&  8&-0.07(10)&-0.05(12)&-0.03(11)&&7.51(10)& 7.53(12)& 7.55(11)\\ % 12 &Mg~I	 7
    &~~~~~II  &&  2&~0.03(01)&&  2&~0.02(00)&-0.01(01)&~0.00(02)&&7.60(00)& 7.57(01)& 7.58(02)\\ %    &~~~~~II
 13 &Al~I     &&  3&~0.14(02)&&  3&~0.04(03)&~0.17(01)&~0.17(03)&&6.51(03)& 6.64(01)& 6.64(03)\\ % 13 &Al~I	 8
 14 &Si~I     && 14&~0.04(10)&& 16&-0.08(19)&~0.05(09)&~0.07(09)&&7.47(19)& 7.60(09)& 7.62(09)\\ % 14 &Si~I	 9
    &~~~~II   &&  2&~0.08(06)&&  3&-0.06(08)&-0.04(09)&-0.01(09)&&7.49(08)& 7.51(09)& 7.54(09)\\ %    &~~~~II
 16 &S~~I     &&  6&~0.15(08)&&  7&~0.02(10)&~0.15(08)&~0.17(07)&&7.35(10)& 7.48(08)& 7.50(07)\\ % 16 &S~~I	10
 19 &K~~I     &&   &~	     &&  1&-0.32    &-0.30    &-0.28	&&4.80	  & 4.82    & 4.84    \\ % 19 &K~~I	11
 20 &Ca~I     && 20&-0.19(12)&& 22&-0.09(11)&-0.17(11)&-0.14(11)&&6.27(11)& 6.19(11)& 6.22(11)\\ % 20 &Ca~I	12
    &~~~~II   &&  2&-0.04(07)&&  3&~0.03(00)&~0.02(10)&~0.01(09)&&6.39(00)& 6.38(10)& 6.37(09)\\ %    &~~~~II
 21 &Sc~II    && 10&~0.09(07)&& 10&-0.03(12)&~0.03(08)&~0.11(08)&&3.14(12)& 3.20(08)& 3.28(08)\\ % 21 &Sc~II	13
 22 &Ti~II    && 34&~0.06(07)&& 38&-0.03(09)&~0.00(09)&~0.07(09)&&4.99(09)& 5.02(09)& 5.09(09)\\ % 22 &Ti~II	14
 23 &V~~I     &&  3&~0.15(03)&&  3&~0.06(02)&~0.16(06)&~0.15(05)&&4.06(02)& 4.16(06)& 4.15(05)\\ % 23 &V~~I	15
    &~~~~II   &&  5&~0.23(12)&&  5&~0.21(14)&~0.14(16)&~0.26(11)&&4.21(14)& 4.14(16)& 4.26(11)\\ %    &~~~~II
 24 &Cr~I     && 22&~0.04(06)&& 22&-0.02(10)&~0.04(05)&~0.08(06)&&5.65(10)& 5.71(05)& 5.75(06)\\ % 24 &Cr~I	16
    &~~~~II   && 23&~0.07(09)&& 26&~0.02(10)&~0.03(10)&~0.08(09)&&5.69(10)& 5.70(10)& 5.75(09)\\ %    &~~~~II
 25 &Mn~I     && 13&~0.09(06)&& 14&-0.04(10)&~0.09(08)&~0.12(05)&&5.35(10)& 5.48(08)& 5.51(05)\\ % 25 &Mn~I	17
 26 &Fe~I     &&   &	     &&127&~0.02(10)&~0.03(08)&~0.07(05)&&7.52(10)& 7.53(10)& 7.57(10)\\ % 26 &Fe~I	18
    &~~~~II   &&   &	     && 31&-0.01(11)&-0.07(11)&-0.01(12)&&7.49(11)& 7.43(11)& 7.49(12)\\ %    &~~~~II
 27 &Co~I     &&  2&~0.27(04)&&  3&~0.14(06)&~0.24(08)&~0.28(07)&&5.06(06)& 5.16(08)& 5.20(07)\\ % 27 &Co~I	19
    &~~~~II   &&   &~	     &&  1&~0.25    &~0.24    &~0.24	&&5.17	  & 5.16    & 5.16    \\ %    &~~~~II
 28 &Ni~I     && 51&~0.21(08)&& 51&~0.18(09)&~0.20(08)&~0.25(09)&&6.43(09)& 6.45(08)& 6.50(09)\\ % 28 &Ni~I	20
 29 &Cu~I     &&  2&~0.25(02)&&  2&~0.25(07)&~0.27(03)&~0.32(01)&&4.46(07)& 4.48(03)& 4.53(01)\\ % 29 &Cu~I	21
 30 &Zn~I     &&  4&~0.35(13)&&  4&~0.22(15)&~0.36(13)&~0.36(15)&&4.82(15)& 4.96(13)& 4.96(15)\\ % 30 &Zn~I	22  Ge 23
 34 &Se~I     &&  1&~0.35    &&  1&~0.23    &~0.34    &~0.39	&&3.64	  & 3.75    & 3.80    \\ % 34 &Se~I	24
 38 &Sr~II    &&   &~	     &&  2&~0.38(10)&~0.33(10)&~0.45(04)&&3.35(10)& 3.30(10)& 3.42(04)\\ % 38 &Sr~II	25
 39 &Y~~II    && 14&~0.66(11)&& 15&~0.58(13)&~0.61(13)&~0.69(14)&&2.82(13)& 2.85(13)& 2.93(14)\\ % 39 &Y~~II	26
 40 &Zr~II    &&  4&~0.37(12)&&  5&~0.36(20)&~0.33(18)&~0.24(15)&&2.96(20)& 2.93(18)& 2.84(15)\\ % 40 &Zr~II	27  Nb, 28 Mo, 29  Ru 30
 56 &Ba~II    &&  2&~0.71(14)&&  3&~0.56(24)&~0.53(20)&~0.49(08)&&2.69(24)& 2.66(20)& 2.62(08)\\ % 56 &Ba~II	31
 57 &La~II    && 11&~0.67(14)&& 18&~0.61(12)&~0.61(10)&~0.64(12)&&1.78(12)& 1.78(10)& 1.81(12)\\ % 57 &La~II	32
 58 &Ce~II    && 20&~0.52(08)&& 31&~0.49(05)&~0.50(10)&~0.56(11)&&2.07(05)& 2.08(10)& 2.14(11)\\ % 58 &Ce~II	33
 59 &Pr~II    &&  2&~0.61(00)&&  6&~0.52(04)&~0.55(05)&~0.63(10)&&1.23(04)& 1.26(05)& 1.34(10)\\ % 59 &Pr~II	34
    &~~~~III  &&   &~	     &&  2&~0.50(02)&~0.50(02)&~0.55(07)&&1.21(02)& 1.21(02)& 1.26(07)\\ %    &~~~~III
 60 &Nd~II    && 15&~0.62(10)&& 32&~0.51(11)&~0.53(10)&~0.58(13)&&2.01(11)& 2.03(10)& 2.08(13)\\ % 60 &Nd~II	35
    &~~~~~III &&   &~	     &&  3&~0.49(02)&~0.50(06)&~0.49(02)&&1.99(02)& 2.00(06)& 1.99(02)\\ %    &~~~~~III
 62 &Sm~II    &&  1&~0.68    &&  8&~0.51(04)&~0.51(12)&~0.55(11)&&1.52(04)& 1.52(12)& 1.56(11)\\ % 62 &Sm~II	36
 63 &Eu~II    &&  2&~0.83(00)&&  4&~0.76(11)&~0.71(09)&~0.81(07)&&1.27(11)& 1.22(09)& 1.32(07)\\ % 63 &Eu~II	37
 64 &Gd~II    &&  1&~0.84    &&  8&~0.97(05)&~0.89(11)&~0.93(11)&&2.09(05)& 2.01(11)& 2.05(11)\\ % 64 &Gd~II	38
 65 &Tb~II    &&   &~	     &&  1&~0.89    &~0.81    &~0.89	&&0.79	  & 0.70    & 0.79    \\ % 65 &Tb~II	39
 66 &Dy~II    &&  2&~0.78(05)&&  7&~0.78(18)&~0.84(10)&~0.87(11)&&1.92(18)& 1.98(10)& 2.01(11)\\ % 66 &Dy~II	40
 68 &Er~II    &&  1&~1.03    &&  8&~0.84(16)&~0.89(18)&~0.93(18)&&1.77(16)& 1.82(18)& 1.86(18)\\ % 68 &Er~II	41
 69 &Tm~II    &&  1&~0.89    &&  1&~0.79    &~0.91    &~0.93	&&0.79	  & 0.91    & 0.93    \\ % 69 &Tm~II	42
 70 &Yb~II    &&   &~	     &&  1&~0.97    &~0.97    &~0.97	&&2.05	  & 2.05    & 2.05    \\ % 70 &Yb~II	43
 71 &Lu~II    &&   &~	     &&  1&~0.99    &~0.96    &~1.06	&&1.05	  & 1.02    & 1.12    \\ % 71 &Lu~II	44
 72 &Hf~II    &&   &~	     &&  1&~0.76    &~0.43    &~0.50	&&1.64	  & 1.31    & 1.38    \\ % 72 &Hf~II	45  Hf,    Ta, 46 Os, 47 Pt 48
 90 &Th~II    &&   &~	     &&  1&~1.02    &~1.02    &~1.04	&&1.11	  & 1.11    & 1.13    \\ % 90 &Th~II	49
\hline
\end{tabular}
\label{meanT}
\end{table}

\begin{table}
\caption { Mean abundance of chemical elements in the atmosphere of \dsct.
	   IUE observations}
\begin{tabular}{rl c ll }
\hline
  Z & Ident. & n& $*$--GS98 &  logN  \\
\hline
  4& Be II &  2  &  -0.29(11)	& 1.06(11)  \\
 32& Ge I  &  2  &  ~0.49(03)	& 3.90(03)  \\
 41& Nb II &  2  &  ~0.47(00)	& 1.89(00)  \\
 42& Mo II &  4  &  ~0.72(24)	& 2.64(24)  \\
 44& Ru II &  1  &  ~0.53	& 2.37	    \\
 72& Hf II &  2  &  ~0.99(01)	& 1.87(01)  \\
 73& Ta II &  1  &  ~1.18	& 1.05	    \\
 76& Os II &  2  &  ~0.92(03)	& 2.37(03)  \\
 78& Pt I  &  1  &  ~0.83	& 2.63	    \\
\hline
\end{tabular}
\label{mIUE}
\end{table}

\end{document}